\documentclass[reprint,prl,twocolumn,superscriptaddress,showpacs,nofootinbib,floatfix,preprintnumbers]{revtex4-1}
\usepackage{graphicx,bm}

\def \PMET{\rm p{\!\!\!/}_T}
\def \MET{\rm E{\!\!\!/}_T}
\begin{document}
\preprint{DAMTP-2014-63} 
\title{Resonant Slepton Production Yields
  CMS $eejj$ and $e \rm p{\!\!\!/}_T jj$ Excesses}

\author{Ben Allanach} \affiliation{DAMTP, CMS, Wilberforce Road, University of
  Cambridge, Cambridge, CB3 0WA, United Kingdom}

\author{Sanjoy Biswas } \affiliation{Dipart.\ di Fisica, Universit\`a di Roma
  “La Sapienza”, Piazzale Aldo Moro 2, I-00185 Rome, Italy}

\author{Subhadeep Mondal } \affiliation{Harish-Chandra Research Institute, Chhatnag Road, Jhusi, Allahabad 211019, India}

\author{Manimala Mitra} \affiliation{Institute for Particle
  Physics Phenomenology, Department of Physics, Durham University, Durham DH1
  3LE, United Kingdom}

\begin{abstract}
Recent CMS searches for di-leptoquark production report local excesses of
2.4$\sigma$ in a $eejj$ channel and 2.6$\sigma$ in a $e \PMET jj$ channel.
Here, we simultaneously explain both 
excesses with resonant slepton production in ${\mathcal R}-$parity violating
supersymmetry (SUSY). We consider resonant slepton
production, which decays to a lepton and a chargino/neutralino, followed by
three-body decays of
the neutralino/chargino via an $\mathcal{R}-$parity violating
coupling. There are regions of parameter space which are also compatible at
the 95$\%$ confidence level (CL) with a 2.8$\sigma$ $eejj$ excess in a recent
CMS  $W_R$ search, while 
being compatible with other direct search constraints. Phase-II of the GERDA
neutrinoless double beta decay ($0\nu\beta\beta$) experiment will probe a
sizeable portion of the good-fit region. 
\end{abstract}
\maketitle

The recent CMS search 
for di-leptoquark production found, with a certain set of cuts, a
2.4$\sigma$ local excess in the $eejj$ channel and a 2.6$\sigma$ local excess
in a $e 
\PMET jj$ channel\footnote{CMS refers to this channel as $e\nu jj$, and we
 shall from here use the same nomenclature.} in comparison to Standard Model
(SM) expectations.  The CMS 
searches use $pp$ 
collision data at the Large Hadron 
Collider (LHC) and a centre of mass energy of $8$ TeV and $19.6 \textrm{fb}^{-1}$
of integrated luminosity. 
Requiring a certain set of cuts (called `$M_{LQ}=650$ GeV' cuts), CMS reported
36 events on a background\footnote{We have added
  systematic and statistical errors in quadrature.} of $20.5 \pm 3.5$ in the
$eejj$ channel, and 18 events on a background of $7.5 \pm 1.6$ in the $e
\nu jj$ channel~\cite{CMS1}. Taken simultaneously and ignoring correlations
between the 
systematics, these excesses amount to a 3.5$\sigma$ effect.
In addition, a $W_R$ search (with different cuts to the di-leptoquark search)
reported a 2.8$\sigma$ excess in the $eejj$ channel at 1.8 TeV$< M_{eejj}<$2.2
  TeV~\cite{Khachatryan:2014dka}. 
These excesses are not significant enough to claim a
discovery, or even evidence. They are similar enough to attempt a unified
explanation of all three, and 
a timely explanation
before the next LHC run (Run II) in terms of new 
physics such that further tests can be applied and analysis strategies can be
set for Run II.

There have been a few attempts to explain the CMS
excesses with  different models. Coloron-assisted
leptoquarks were proposed in Ref.~\cite{Bai:2014xba}. 
The $W_R$ excess was interpreted in GUT models in
Refs.~\cite{Deppisch:2014qpa,Heikinheimo:2014tba}. In
Ref.~\cite{Dobrescu:2014esa}, pair production of vector-like leptons
was proposed via $W'/Z'$ vector bosons.
Ref.~\cite{Aguilar-Saavedra:2014ola} performed a detailed analysis (including a
general flavor structure) of $W'/Z'$ interpretations of the $W_R$ search data.
In ref.~\cite{Queiroz:2014pra}, it was supposed that leptoquarks consistent with
the di-leptoquark excess decay into dark matter particles with a significant
branching ratio. 
Ref.~\cite{Chun:2014jha} explains the di-leptoquark excesses with di-sbottom
production, followed by ${\mathcal R-}$parity violating (RPV) decay.
In a previous letter~\cite{Allanach:2014lca}, we proposed that resonant
slepton production was 
responsible for the $W_R$ search excess in RPV supersymmetry.
One of us showed that this explanation is also consistent with recent
deviations from SM prediction measured by LHCb in $B^+ \rightarrow K^+ ll$
decays~\cite{Biswas:2014gga}.

\begin{figure}[ht]
 \includegraphics[width=7.5cm]{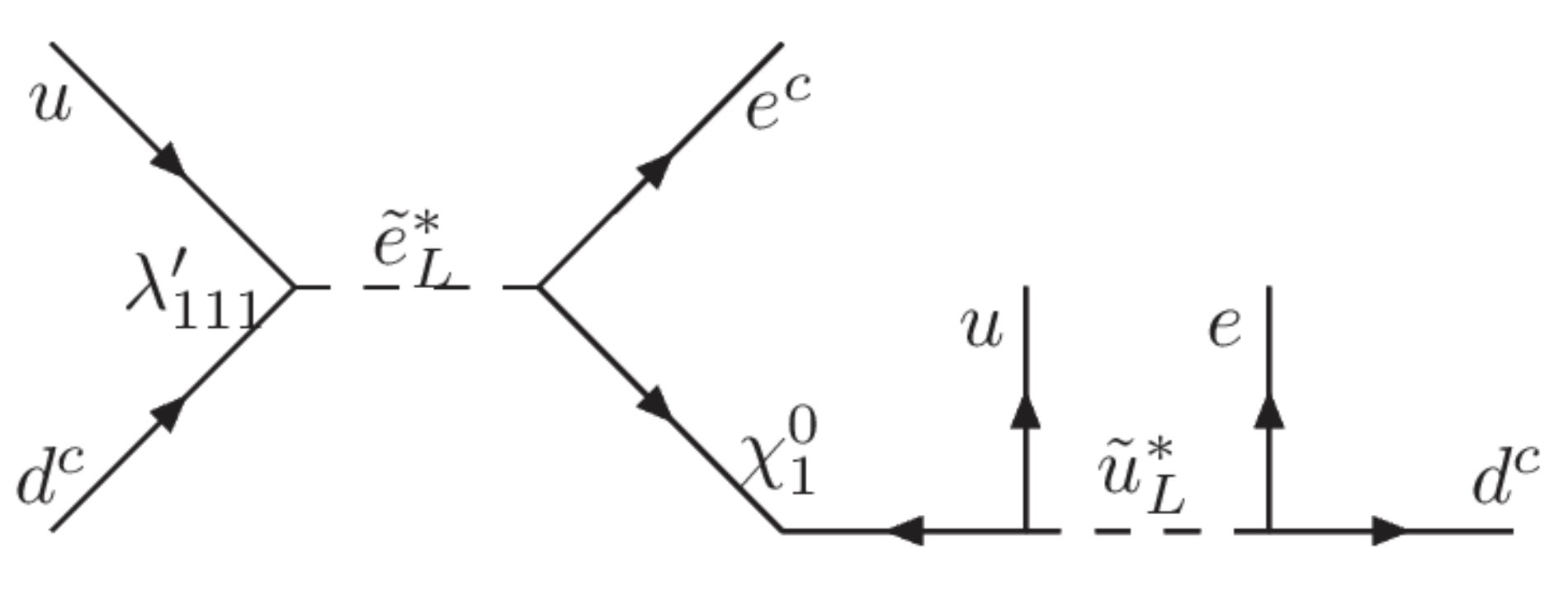}
 \includegraphics[width=7.5cm]{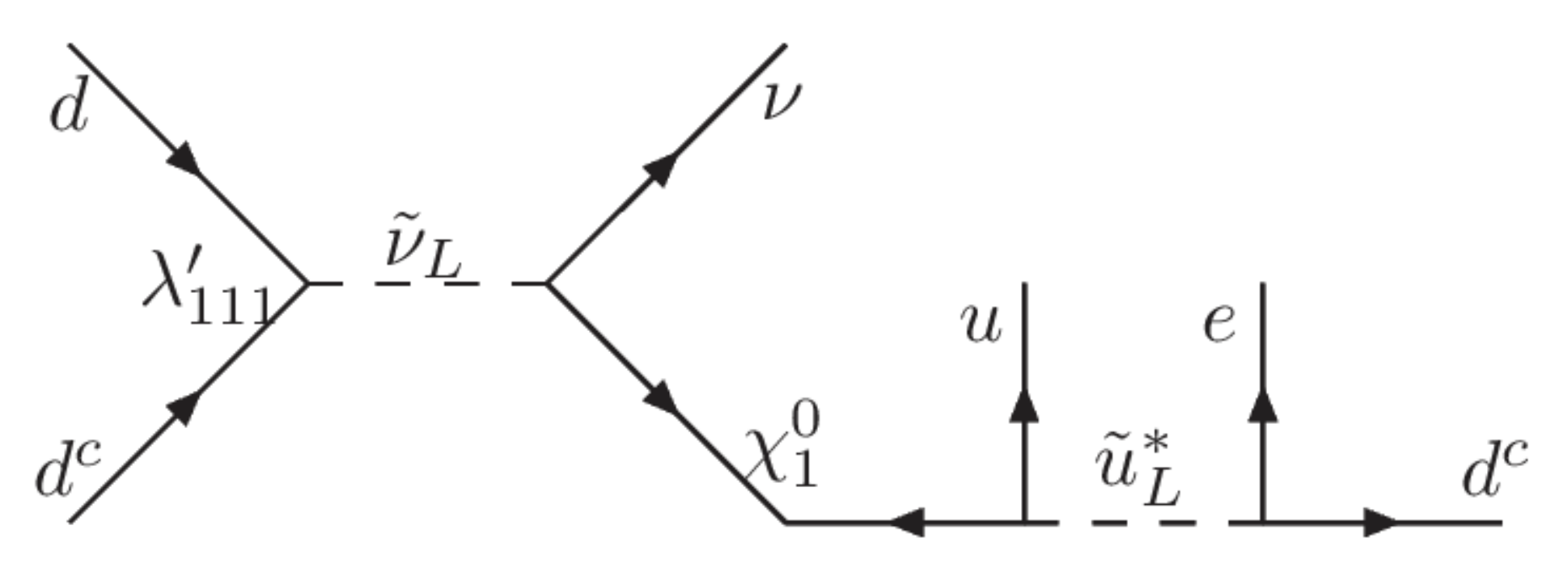}
 \caption{Feynman diagrams for single slepton production leading to a
   $eejj(e \nu jj)$ signal at the LHC\@. Other diagrams, where the $\chi_1^0$ is
   replaced by $\chi_1^\pm$ (among other replacements) also contribute.}
 \label{fig:FDsel}
\end{figure}

In the present letter, we shall show that RPV resonant slepton production can
simultaneously 
fit the two excesses in the di-leptoquark search while remaining consistent
with other direct searches, including the $W_R$ search data. 
$\mathcal{R}$-parity
is a multiplicative discrete symmetry
defined as 
$\mathcal{R}=(-1)^{3(B-L)+2S}$, where $B$ and $L$ correspond to baryon and
lepton number, and $S$ is the spin.  
In particular, we show that
RPV with a non-zero $\lambda'_{111}$ coupling
can fit the 
CMS excesses \cite{Khachatryan:2014dka,CMS1} via resonant slepton
production (with a left-handed slepton mass of around $m_{\tilde l} \sim 2$
TeV) in $pp$ collisions.   
The slepton (either a selectron or an electron sneutrino) then decays, as
shown in Fig.~\ref{fig:FDsel}, either into $eejj$ or $e \nu
jj$. 
For sleptons much heavier than $M_Z$, one automatically\footnote{Charged
  slepton production is larger because it couples to $u$ in the proton
  rather than the $d$.} gets
cross-sections of the same order of magnitude for the $eejj$ and the $e\nu jj$
channels
because their 
masses are at tree-level related by~\cite{softsusy} 
\begin{equation}
m_{\tilde e_L}^2 = m_{\tilde \nu_L}^2 + M_W^2 \cos 2 \beta,
\end{equation}
where $\cos 2\beta<0$, $\tan \beta$ is the ratio of the two MSSM Higgs
doublet vacuum expectation values, and we have neglected small terms
proportional to powers of lepton masses. For $m_{\tilde e_L}$ around 2 TeV 
(which we shall be interested in), $m_{\tilde e_L} \approx m_{\tilde \nu_L}$
is a good approximation. In addition, the parton distribution functions for
anti-up quarks $u^c$ and anti-down quarks $d^c$ are similar within the proton,
resulting in cross-sections for the two processes shown in
Fig.~\ref{fig:FDsel} of a similar order of magnitude. 

The $\lambda'_{111}$ term in the RPV superpotential is\footnote{$^c$ denotes
  the charge conjugate.} 
\begin{equation}
  W_{\not{R}}= \lambda'_{111} LQd^c. 
  \label{eqrpv}
\end{equation}
This induces the following Lagrangian terms,
\begin{equation}
  \mathcal{L}=\lambda'_{111} \left(-\tilde{e} u d^c- \tilde{u} e d^c+
\tilde{d} \nu_e d^c+ \tilde{\nu}_e d d^c+\ldots\right) \\ \label{eq:L}
\end{equation}
The $\lambda'_{111}$ coupling in Eq.~\ref{eq:L} can lead to  single slepton
production at hadron colliders, as 
first studied in \cite{Dimopoulos:1988fr} and subsequently in 
\cite{Hewett:1998fu, Dreiner:1998gz, Dreiner:2000vf, Dreiner:2000qf,
  Dreiner:2012np, Allanach:2009xx, Allanach:2009iv}. For an example slepton 
mass of $m_{\tilde l}=2.1$ TeV and $0.03<\lambda'_{111}<0.5$ the
production cross-section varies from less than 1 fb to as high as 130
fb \cite{Dreiner:2000vf}.  

The coupling $\lambda'_{111}$ responsible for single slepton production in
Fig.~\ref{fig:FDsel} induces
$0\nu \beta \beta$ \cite{0nu2beta-old,
  ms0nu2beta,  hirsch}, which is  
 not permitted in the SM because of its prediction of lepton number
 conservation. 
 The present bound on the $0\nu\beta\beta$ half-life  of $^{76}\textrm{Ge}$
 is 
  $T^{0\nu}_{1/2}>2.1 \times 10^{25}\, \textrm{yrs}$ at 90$\%$CL 
from GERDA~\cite{gerda},
  while the 90$\%$ CL combined bound on the half-life  from  previous
  experiments 
  is $T^{0\nu}_{1/2}>3.0 \times 
  10^{25}\, \textrm{yrs}$~\cite{gerda}. The future  
$0\nu \beta \beta$ experiment  GERDA Phase-II will be commissioned soon and
is expected to  
improve the half-life sensitivity to  $T^{0\nu}_{1/2} \sim 2\times 10^{26} \,
\textrm{yrs}$~\cite{gerdafuture}.
A positive signal in $0\nu \beta \beta$ experiments is likely to be
interpreted in terms of a Majorana nature of the light neutrinos, 
but instead it could be  in part, or dominantly, due to RPV SUSY\@. 
There are several contributing diagrams including slepton,
neutralino, squark 
and/or gluino exchange, but for high squark and gluino masses, the dominant one
often involves 
internal sleptons and lightest neutralinos $\chi_1^0$~\cite{Allanach:2014lca}. 
As pointed out in Ref.~\cite{Allanach:2009iv}, one can then marry resonant
slepton search data from the LHC with the predicted $0\nu\beta\beta$ rate in
order to provide further tests and interpretations. We shall here neglect
 contributions to $0\nu\beta\beta$ coming from neutrino masses, assuming the
 one due to RPV is  dominant. 

It is our aim to see if resonant slepton production and decay can fit
the CMS di-leptoquark excesses while evading other experimental constraints, and
to examine the compatibility with our previous resonant 
slepton explanation of the $W_R$ excess. 
Then, we wish to explore the $0\nu\beta\beta$ decay experiments' prospects
within any good-fit region.
In fact, the strongest indirect bound that is relevant to our
analysis is that from $0\nu\beta\beta$.
Other indirect bounds
on the $\lambda'_{111}$ coupling can be 
found in Ref.~\cite{Allanach:1999ic,Reviews}. 
For example, the RPV violating contribution to the anomalous magnetic
moment of 
the electron $g_e$ is\footnote{In order to obtain this formula, we 
converted
  the approximate expression in Ref.~\protect\cite{Bhattacharyya:2009hb} for the anomalous  magnetic moment of the 
muon to that of the electron.}
\begin{equation}
\delta \frac{g_e-2}{2} \sim
\frac{|\lambda'_{111}|^2 m_e^2}{32 \pi^2 \tilde m^2},
\end{equation}
where $\tilde m$ is the size of supersymmetric particle masses appearing in
the one-loop diagram. Putting $\lambda'_{111}=1$ and $\tilde m=1$ TeV, we
obtain a contribution of $< 10^{-15}$, far below current
bounds: the difference between the experimental value and the
Standard Model prediction is~\cite{Aoyama:2012wj} $(-1.06 \pm 0.82) \times 10^{-12}$. 

We shall follow  a bottom-up phenomenological 
approach. We decouple sparticles which
are not relevant for our hypothesised signals. Otherwise, we  
fix the first generation lightest neutralino mass $M_{\chi_1^0}$ to be 
1 TeV (although we have checked that there are only small deviations in our
predictions and constraints if we reduce this to 0.8 TeV)
the slepton mass varies from 1.8 TeV up to 2.2 TeV and all other
sparticles are above the TeV scale. 
The squark and gluino masses are fixed at 2.5 TeV.
We  set other RPV couplings to zero, allowing us to focus purely  on the
effects of $\lambda'_{111}$. 

We have 
considered the following representative scenarios:\\
{\bf S1:} $M_1 < M_2=M_1+200 < \mu$, {i.e.}, the LSP is mostly bino-like
  with a small wino-component.  In this case the slepton has a substantial
  branching ratio of decays to the second lightest neutralino or lightest
  chargino $\chi_1^\pm$.\\
{\bf S2:} 
$M_1 < \mu < M_2 $, the LSP is still dominated by the
  bino-component, with a heavy intermediate higgsino mass and an even heavier
  wino mass ($> 1$ TeV). This case increases the
  branching ratio of slepton decays into the lightest neutralino and a
  lepton compared to {\bf S1}.\\
{\bf S3:} $M_2 \ll M_1 \simeq \mu$, {i.e.} the LSP is dominantly
  wino-like. In this case,  slepton  decay to $\chi_1^\pm$ and $\chi_1^0$
 with a substantial branching fraction, which then subsequently
 decay via $\lambda'_{111}$. 
\begin{table}
\begin{tabular}{|c|ccccc|} \hline
$\sigma_{95}$/fb & 160 & 75 & 50 & 45 & 36 \\
$m_{\tilde l}$/TeV &1.8 & 1.9 & 2.0 & 2.1 & 2.2 \\ \hline
\end{tabular}
\caption{95$\%$ upper bound on cross-section times branching ratio times
  acceptance for resonantly produced sleptons decaying to
  di-jets~\cite{Chatrchyan:2013qha}.} 
\label{tab:lim}
\end{table}
Depending on the nature of the lightest
neutralino and the value of the $\lambda'_{111}$ coupling, the branching ratio
changes considerably \cite{Dreiner:2012np}. At large values of
$\lambda'_{111}$, the branching ratio of a slepton into two jets becomes larger.
Thus, resonant slepton production then becomes constrained by
 di-jet resonance searches~\cite{Dreiner:2012np}. 
We take into account the constraint from a CMS di-jet resonance
search \cite{Chatrchyan:2013qha}: the upper limits are displayed in
Table~\ref{tab:lim}.

\begin{figure}[t]
  \includegraphics[width=4.2cm]{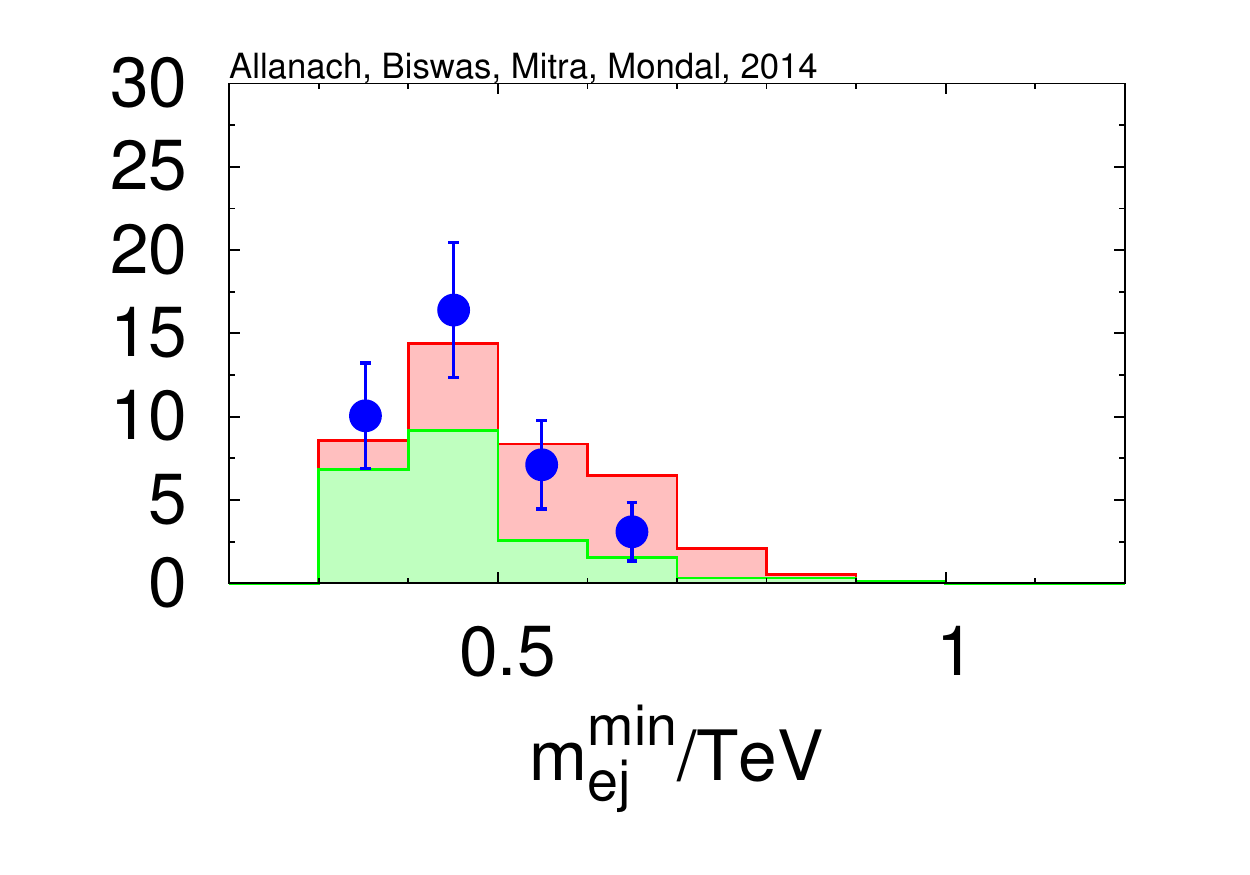}
  \includegraphics[width=4.2cm]{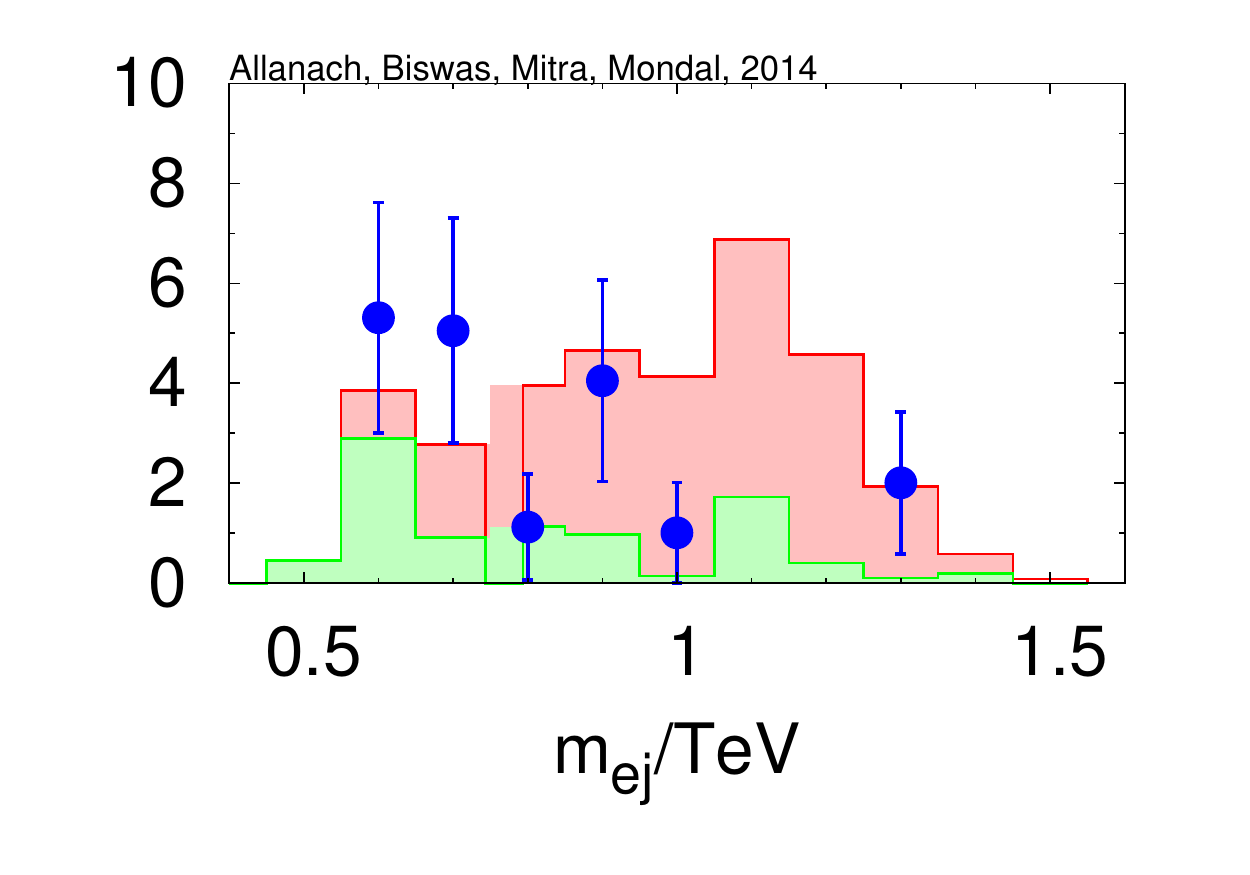}
  \includegraphics[width=4.2cm]{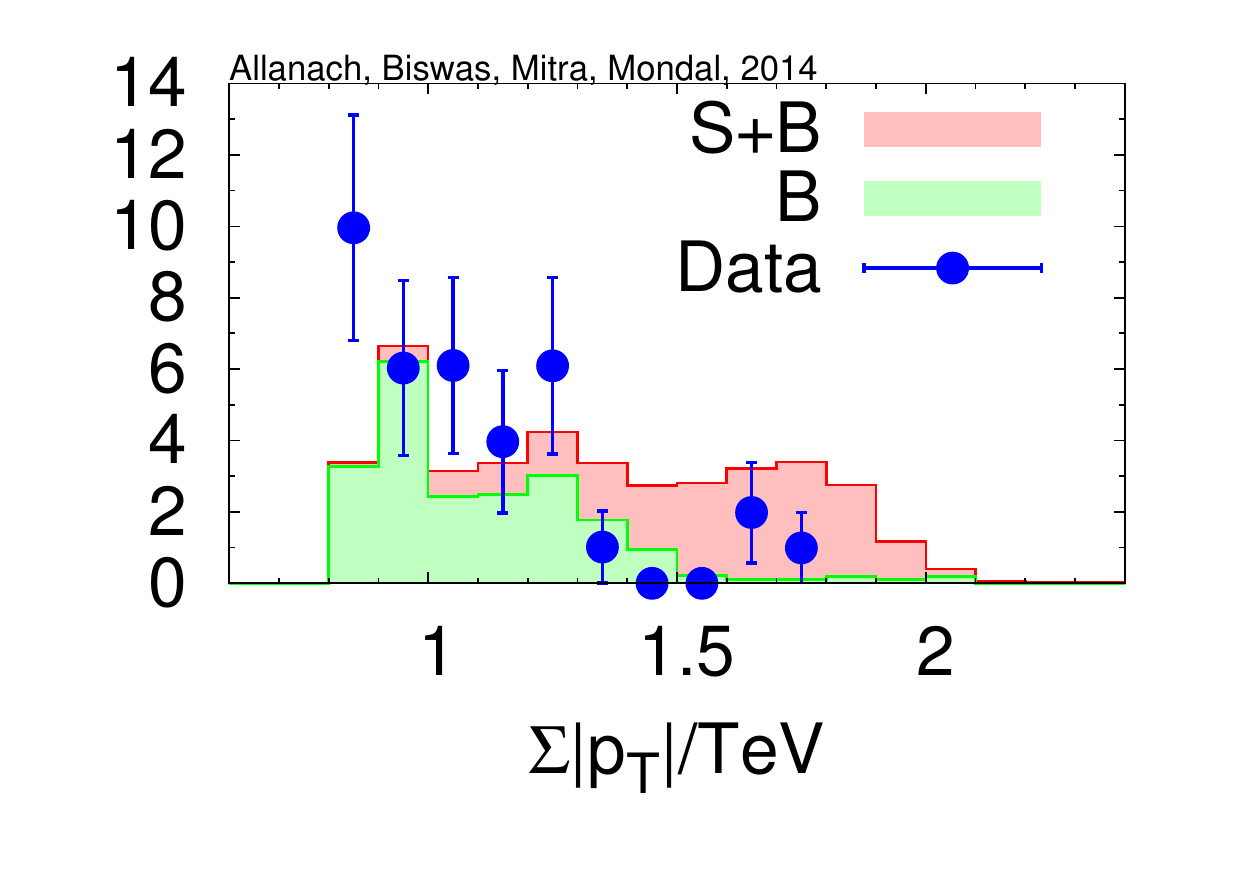}
  \includegraphics[width=4.2cm]{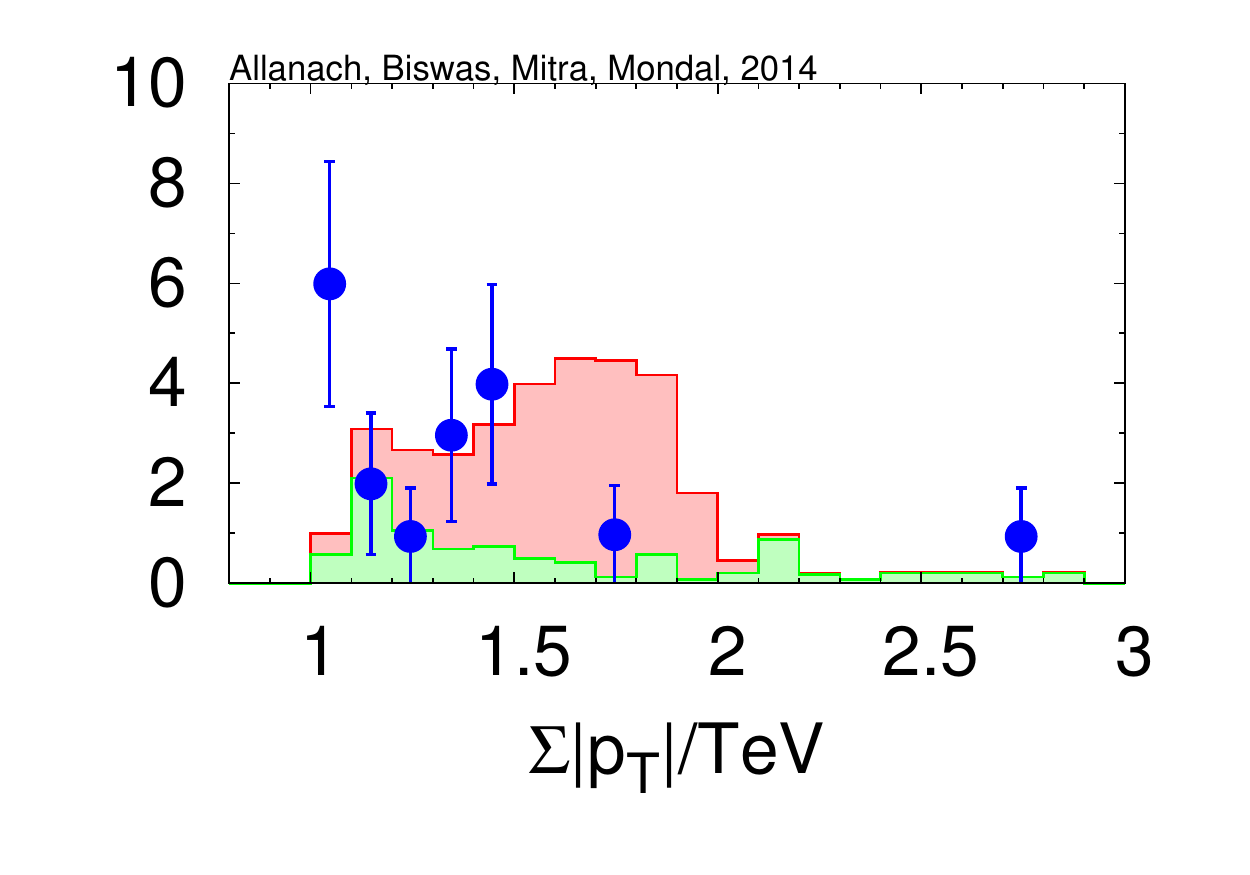}
  \vspace*{-0.5cm}  
  \caption{{A comparison of the measured (`Data'), background (`B') and
      example signal plus background (`S+B')
       distributions for $m_{ej}^{min}(eejj)$, $m_{ej}(e \nu jj)$
       and the scalar sum of the $p_T$s of visible objects for (lower left)
       $eejj$ and (lower right) $e \nu jj$. 
       $m_{ej}^{min}$ is the invariant mass of 
  the electron-jet pairing
  combination for which the {\em difference}\/ between the $m_{ej}$ for each
  pair is   smallest. The vertical axis measures the
       number of events 
       falling in a bin
       after imposing cuts. 
       The signal point corresponds to
       $\lambda'_{111}=0.175$, $m_{\tilde l}=2$ TeV and $M_{\chi_1^0}=0.9$ TeV ({\bf
        S2}). Data and SM backgrounds are taken from
      \cite{CMS1}.
}}
  \label{fig:mej}
\end{figure}
We simulate first generation resonant slepton production in $pp$
collisions at a centre of mass energy $\sqrt{s}=8$ TeV using CalcHEP (v3.4.2)
\cite{Belyaev:2012qa},
and the subsequent decay, showering and hadronization effects have been
performed 
by PYTHIA (v6.4) \cite{Sjostrand:2006za}. We use SARAH-v4.0.1 \cite{sarah} and SPheno-v3.2.4 \cite{spheno}
for the model implementation and to compute  branching ratios. We approximate
the next-to-leading order QCD corrections by multiplying the tree-level
production cross section 
with a $K-$factor of 1.34~\cite{Dreiner:2012np}. We use CTEQ6L parton
distribution
functions~\cite{Kretzer:2003it}  with factorization and  renormalization
scales set at 
$m_{\tilde l}$. To take the detector resolution into account, we  also use various
resolution functions 
parameterized as in \cite{Chatrchyan:2011ds} for the final state objects.

The final states studied in \cite{CMS1}, contain either exactly two
isolated $e$s and at least two jets ($eejj$), or
one isolated $e$, at least two jets and  missing transverse momentum ($l\nu
jj)$. Basic object 
definitions for the leptons and jets together with final
selection cuts, as 
outlined in \cite{CMS1}, have been imposed as shown
in Table~\ref{tab:Cuts}.
\begin{table}
\begin{tabular}{|c|c|}\hline
Channel & Cuts ($M_{LQ}=650$ GeV)\\ \hline
$eejj$ & $S_T>850$ GeV, $m_{ee}>155$ GeV, $m_{ej}^{min}>360$ GeV. \\
$e \nu jj$ &  $S_T>1040$ GeV, $\MET>145$ GeV, \\
 & $m_{ej}>555$ GeV, $m_T(e
\nu)>270$ GeV \\ \hline
Channel & Cuts ($M_{LQ}=700$ GeV)\\ \hline
$eejj$ & $S_T>1120$ GeV, $m_{ee}>160$ GeV, $m_{ej}^{min}>390$ GeV. \\
$e \nu jj$ &  $S_T>1120$ GeV, $\MET>155$ GeV, \\
 & $m_{ej}>600$ GeV, $m_T(e
\nu)>280$ GeV \\ \hline
Channel & Cuts \\ \hline
$W_R$ & $m_{ee}>200$ GeV, $M_{eejj}>600$ GeV, $2e+\ge2j$ \\
\hline\end{tabular}
\caption{Cuts in each channel, from
  Refs.~\cite{CMS1,Khachatryan:2014dka}. $S_T=\sum | p_T|$ 
  is the scalar sum of transverse momenta of all visible objects, $m_{ee}$ is
  the invariant mass of 
  the lepton pair and    $m_{ej}^{min}$ is the invariant mass of 
  the electron-jet pairing
  combination for which the {\em difference}\/ between the $m_{ej}$ for each
  pair is   smallest. $m_T(e\nu)$ is the electron-neutrino transverse mass,
  and $m_{ej}$ is the electron-jet invariant mass where the lepton is paired
  with the jet that results in the most smallest difference between $m_{ej}$
  and $m_T(e\nu)$. 
} \label{tab:Cuts} 
\end{table}
In their analysis, CMS defined many signal regions, each with its own set of
cuts. We pick one of 
them (designed for 700 GeV di-leptoquark sensitivity) which only shows a small 
excess, to check that our model is not ruled out by it. 

We assume a truncated Gaussian for the prior probability density function
(PDF) of $\bar b \pm \sigma_b$ background events: 
\begin{equation}
p(b | \bar b,\ \sigma_b) = \left\{ 
\begin{array}{lr}
B e^{-(b-\bar b)^2/(2 \sigma_b^2)}& \forall b > 0, \\
0& \forall b \leq 0,
\end{array} \right.
\end{equation}
where $B$ is a normalisation factor that makes the distribution integrate to 1.
We marginalise the Poissonian probability of measuring $n$ events over $b$
in
order to obtain confidence limits:
\begin{equation}
P(n|n_{exp},\ \bar b,\ \sigma_b) = \int_0^\infty db\ p(b | \bar b,\ \sigma_b)
\frac{e^{-n_{exp}} n_{exp}^{n}}{n!}, \label{pois}
\end{equation}
where $n_{exp}$ is the number of expected events.
The CL of $n_{obs}$ observed events is then $P(n\leq n_{obs})$.
Calculated in this way, the $e \nu jj$ excess is a 2.9$\sigma$ effect, and
the $eejj$ excess is a 2.6$\sigma$ effect, making the two combined 3.9$\sigma$.
With a two-tailed 95$\%$ CL, the number of signal events in the $eejj$ channel
 is $s_{eejj} \in [19.4,\ 58.4 ]$, whereas in the $e \nu jj$ channel it is
$s_{e \nu jj} \in [ 7.8,\ 34.3 ]$.
For the $W_R$ search, we combine the statistics from the bins
$M_{eejj}/\text{TeV} \in [1.6-1.8,\ 1.8-2.2,\
2.2-4]$ (bin $i=1,2,3$, respectively). CMS only observed a large excess in the
1.8-2.2 TeV 
bin, there was no large excess in the adjacent
bins and so these help constrain parameter space. Bin $i$ has a $\chi^2$
statistic $\chi^2_i=-2 \ln P(n^{(i)}|n_{exp}^{(i)},\ \bar b^{(i)},
\sigma_b^{(i)})$, where $P$ is obtained from 
Eq.~\ref{pois}.
When considering $W_R$ search
constraints, we therefore form a total $\chi^2=\chi_1^2+\chi_2^2+\chi_3^2$.
Imposing a 95$\%$CL limit is equivalent to limiting the total $\chi^2<3.84$. 

\begin{figure}[!]
  \includegraphics[width=8.5cm]{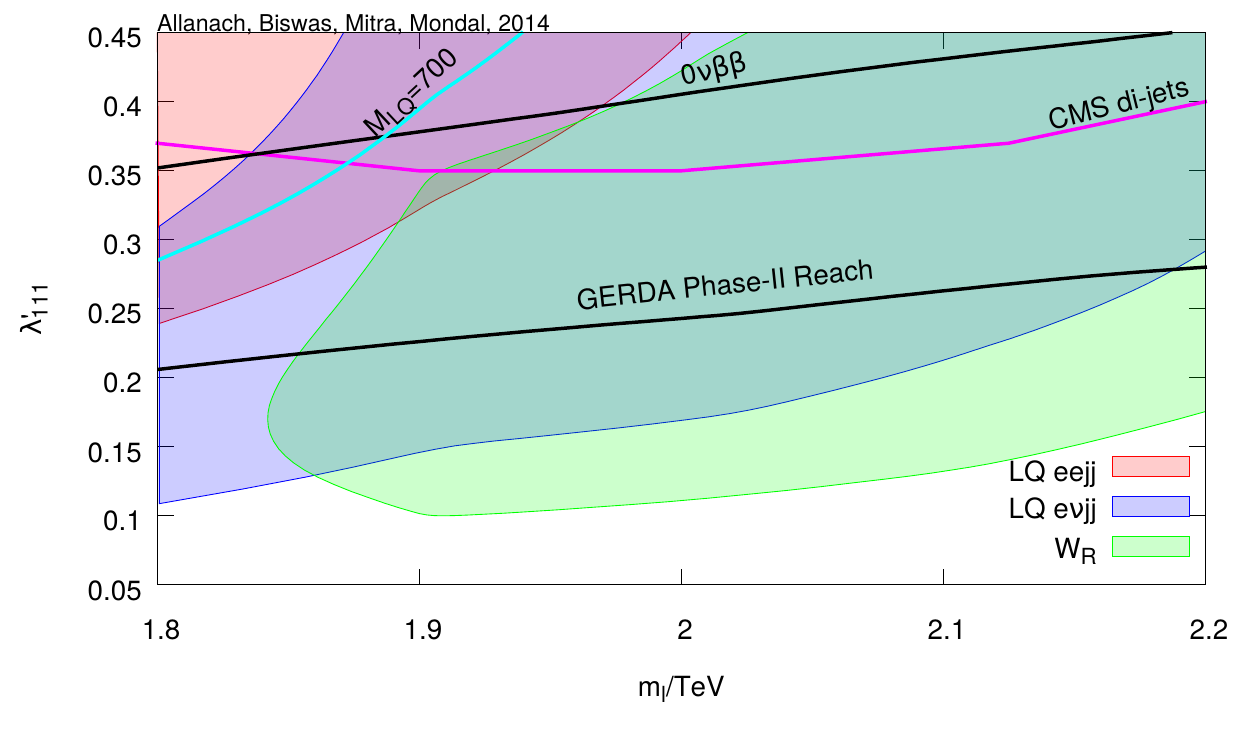}
  \includegraphics[width=8.5cm]{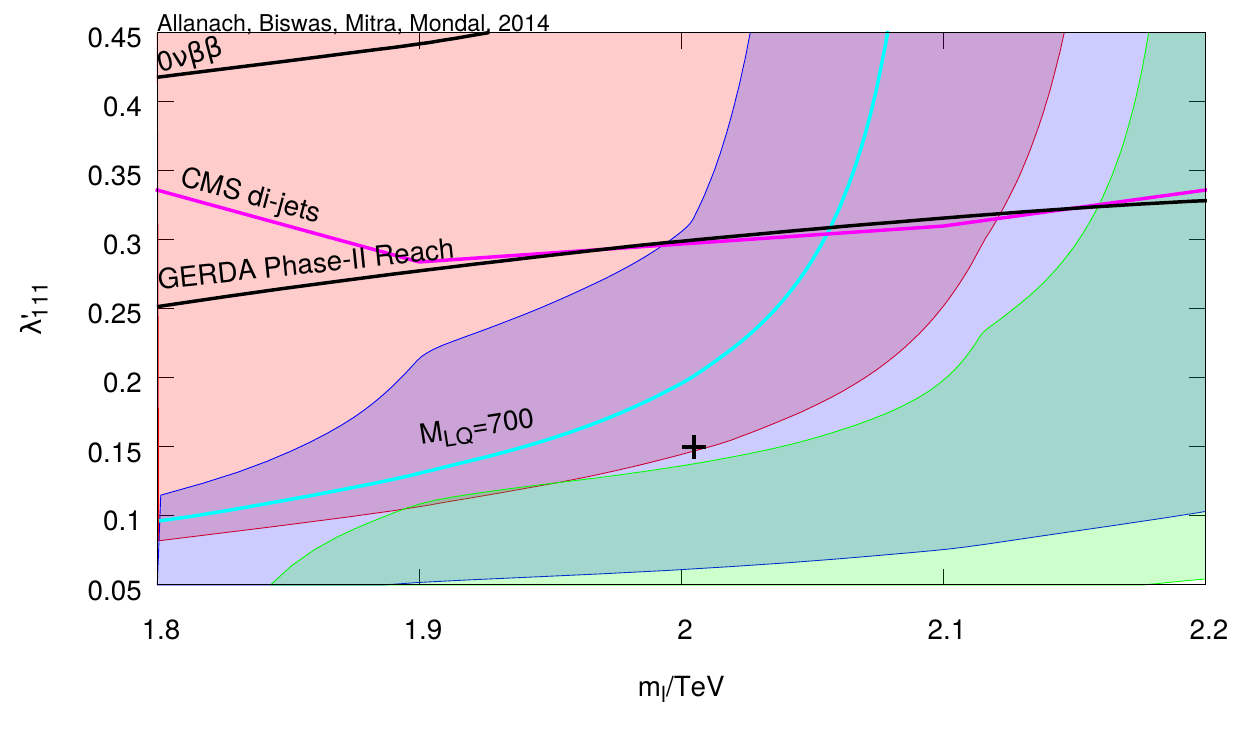}
  \includegraphics[width=8.5cm]{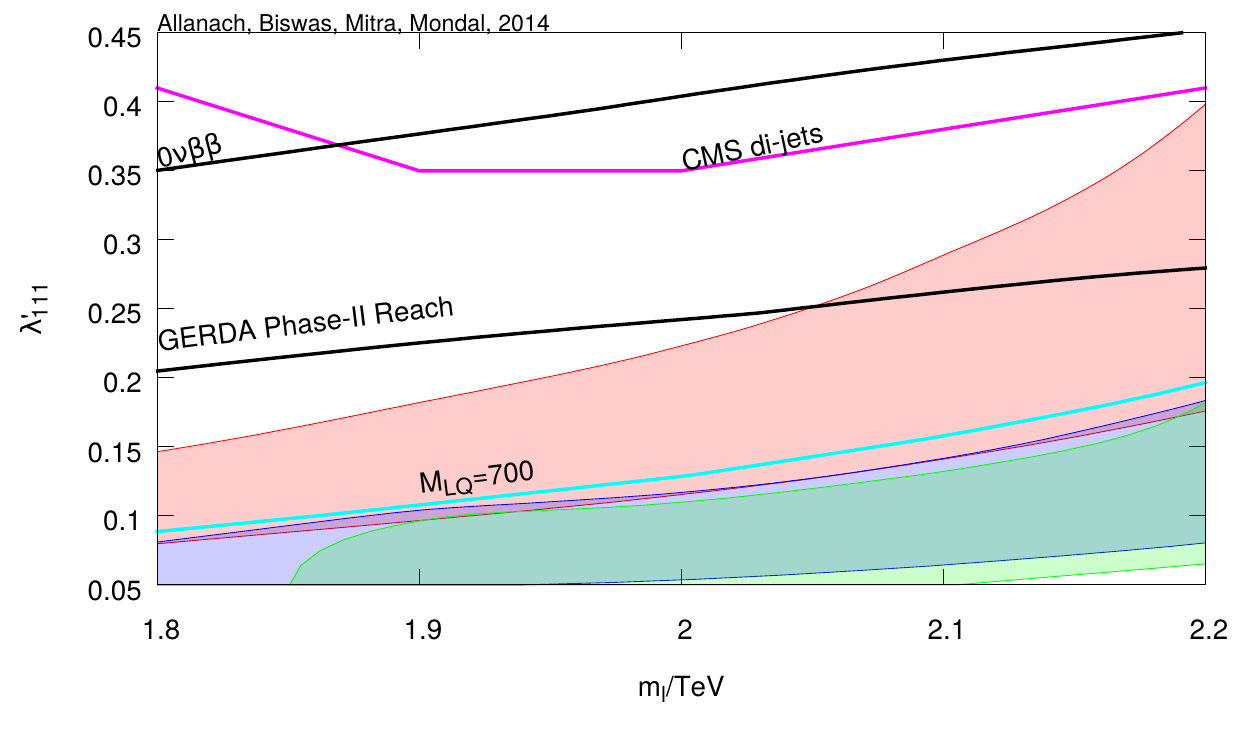}
\vspace*{-0.5cm}  
\caption{Constraints  in the $m_{\tilde l}-\lambda'_{111}$ plane
  assuming $M_{\chi_1^0}=0.9$ TeV in scenario (top)
    {\bf S1}, (middle) {\bf S2} and (bottom) {\bf S3},
    where the nuclear matrix elements have been adopted from 
    \cite{Allanach:2009xx}. The parameter space fitting the $eejj$, $e\nu jj$
    di-leptoquark search excesses and the $W_R$ $eejj$ excess to 95$\%$CL is
    shown in the key. 
    The region above the lines (except for the one labelled `GERDA Phase-II
    Search' is excluded,
    labeled near the edge of exclusion: either the 90$\%$CL constraint from
    current combined neutrinoless double beta decay bounds ($0\nu\beta\beta$)
    or at the 95$\%$CL from the CMS di-jets shape analysis. The expected
    90$\%$ CL exclusion reach from GERDA Phase-II \cite{gerdafuture} is shown
    as the region above the solid black line. The cross in the middle panel
    shows 
    the location of our example point.} 
    \label{fig:scan1}
\end{figure}
We present the predicted event numbers in  Table.~\ref{tab:events} and
Fig.~\ref{fig:mej} for an example point in our parameter space and
an integrated
luminosity of 19.6 fb$^{-1}$. 
\begin{table}[!]
  \small
  \begin{tabular}{|c|c|c|c|} 
    \hline 
    Channel & $s+\bar b$ & $\bar b \pm \sigma_b$ & Data \\
    \hline
    $eejj (M_{LQ}=650$ GeV$)$ & 41.5 & 20.5$\pm$3.3 & 36 \\
    $e \nu jj (M_{LQ}=650$ GeV$)$ & 33.9 & 7.5$\pm$1.6 & 18 \\
    $eejj (M_{LQ}=700$ GeV$)$ & 32.7 & 12.7$\pm$2.7 & 17  \\
    $W_R (1.6<M_{eejj}/\textrm{TeV}<1.8)$& 12.4 & 9.6$\pm$3.8& 10\\
    $W_R (1.8<M_{eejj}/\textrm{TeV}<2.2)$& 26.0 & 4.0$\pm$1.0& 14\\
    $W_R (M_{eejj}/\textrm{TeV}>2.2)$& 2.6 & 2.2$\pm$1.8& 4\\
    \hline
  \end{tabular}
  \caption{Number of events from signal plus central background $s+\bar b$,
    background $b$, background uncertainty $\sigma_b$ and 
 reconstructed data  
 after  
    application of the selection cuts for 19.6 fb$^{-1}$ integrated
    luminosity and 
    8 TeV center of mass energy. The signal model is scenario {\bf S2} with
    $\lambda'_{111}=0.175$,
 $m_{\tilde{l}}=2$ TeV and $M_{\chi_1^0}=0.9$ TeV. The data and SM backgrounds are taken from
 Ref.~\cite{CMS1} and Ref.~\cite{Khachatryan:2014dka}.}
  \label{tab:events}
\end{table}
In Fig.~\ref{fig:mej}, the distributions $m_{ej}^{min}$, $m_{ej}$ and the
scalar sum of $p_T$ for the visible objects are shown for
data~\cite{CMS1},
background and an example signal model point prediction\footnote{The
  kinmetic distributions change very little if the slepton mass is varied by 0.1
  TeV, whereas the normalisation is sensitive to the overall value of
  $\lambda'_{111}$.}. The figure shows that
the distributions are reproduced to a reasonably good level by our model
point, although the model point is perhaps slightly broader and of higher
 energy,
compared to the measurements. One must bear in mind though that the
statistical power of the kinematical distributions is very limited because of
the small statistics.

Fig.~\ref{fig:scan1} shows the
$\lambda'_{111}-m_{\tilde l_L}$ plane for {\bf
  S1}-{\bf S3}.
It is evident that  the largest values of $\lambda'_{111}$ 
that we take are ruled out by the CMS di-jet search
\cite{Chatrchyan:2013qha}. 
In {\bf S1} and {\bf S2}, there is a small region where all constraints are
respected 
and the three excesses are within their 95$\%$CL constraints. In {\bf S1}, the
whole of this region around $\lambda'_{111}=0.32$ and $m_{\tilde l}=1.88$ TeV can
be covered by the GERDA Phase-II~\cite{gerdafuture} 
$0\nu\beta\beta$ 
search, whereas in {\bf S2}, it will not cover the overlap
region (around $1.9<m_{\tilde l}/\textrm{TeV}<2$ and
$0.11<\lambda'_{111}<0.13$): indeed, much of the parameter space that will be
probed by GERDA 
Phase-II is already disfavored by the CMS di-jets search. 
The overlap regions
we have found in {\bf S1} and {\bf S2} are somewhat 
marginal, being on the edge of exclusion for all three excess channels. 
A more thorough search through parameter
space (for example considering higher values of $M_{\chi_1^0}$) might make the
fit better. 
There are however large regions where the two di-leptoquark
channels both fit data well while respecting other constraints. They are
consistent with there being a downward fluctuation in the $W_R$ excess. 
In {\bf S3}, there is no parameter space where the two di-leptoquark channels
give the correct rates, to within a 95$\%$ CL\@. This is because, in {\bf S3},
the chargino has a similar mass to the $\chi_1^0$ and so contributes
significantly, producing higher rates for $e\nu jj$ compared to $eejj$. 
Ideally, one would perform a combined fit between the excesses, taking
into account other measurements. However, this is
precluded by the fact that
the $W_R$ search and the $eejj$ channels contain common
background events, and we do not have access to the correlations between them. 
For the current paper, we content ourselves with a depiction of where the
preferred regions for each measurement lie.

To summarize, the CMS di-leptoquark search excesses are well described by the
hypothesis of resonant slepton production. Both the $eejj$ channels and the
$e\nu jj$ channel event rates can fit the excess well, and the most important
kinematic 
distributions appear by eye to be reasonable. We have found regions parameter
space that are consistent with the 95$\%$CL regions for the di-leptoquark
channels {\em and}\/ the $W_R$ search. 

{\bf Acknowledgements:} This work has been partially supported by STFC\@. 
We thank the Cambridge SUSY working group for stimulating discussions.
MM  acknowledges  partial support of the ITN INVISIBLES 
(Marie Curie Actions, PITN-GA-2011-289442).


\begin{thebibliography}{99}


\bibitem{CMS1}
CMS Collaboration, 
Tech. Rep. CMS-PAS-EXO-12-041,
CERN, Geneva, 2014.

\bibitem{Khachatryan:2014dka} 
  V.~Khachatryan {\it et al.}  [CMS Collaboration],
  arXiv:1407.3683 [hep-ex].

\bibitem{Bai:2014xba} 
  Y.~Bai and J.~Berger,
  arXiv:1407.4466 [hep-ph].

\bibitem{Deppisch:2014qpa} 
  F.~F.~Deppisch, T.~E.~Gonzalo, S.~Patra, N.~Sahu and U.~Sarkar,
  arXiv:1407.5384 [hep-ph].

\bibitem{Heikinheimo:2014tba} 
  M.~Heikinheimo, M.~Raidal and C.~Spethmann,
  arXiv:1407.6908 [hep-ph].

\bibitem{Dobrescu:2014esa} 
  B.~A.~Dobrescu and A.~Martin,
  arXiv:1408.1082 [hep-ph].

\bibitem{Aguilar-Saavedra:2014ola} 
  J.~A.~Aguilar-Saavedra and F.~R.~Joaquim,
  arXiv:1408.2456 [hep-ph].

\bibitem{Queiroz:2014pra} 
  F.~S.~Queiroz, K.~Sinha and A.~Strumia,
  arXiv:1409.6301 [hep-ph].

\cite{Chun:2014jha}
\bibitem{Chun:2014jha} 
  E.~J.~Chun, S.~Jung, H.~M.~Lee and S.~C.~Park,
  arXiv:1408.4508 [hep-ph].

\bibitem{Allanach:2014lca} 
  B.~Allanach, S.~Biswas, S.~Mondal and M.~Mitra,
  arXiv:1408.5439 [hep-ph].

\bibitem{Biswas:2014gga} 
  S.~Biswas, D.~Chowdhury, S.~Han and S.~J.~Lee,
  arXiv:1409.0882 [hep-ph].

\bibitem{softsusy}
 B.~C.~Allanach,
  Comput.\ Phys.\ Commun.\  {\bf 143}, 305 (2002)
  [hep-ph/0104145].

\bibitem{0nu2beta-old}
 G.~Racah,
  Nuovo Cim.\  {\bf 14}, 322-328 (1937);
  W.~H.~Furry,
  Phys.\ Rev.\  {\bf 56}, 1184-1193 (1939).
\bibitem{ms0nu2beta}
 R.~N.~Mohapatra,
 Phys.\ Rev.\  {\bf D34}, 3457-3461 (1986);
%
 K.~S.~Babu, R.~N.~Mohapatra,
  Phys.\ Rev.\ Lett.\  {\bf 75}, 2276-2279 (1995)
  [hep-ph/9506354].
%
\bibitem{hirsch}
  M.~Hirsch, H.~V.~Klapdor-Kleingrothaus, S.~G.~Kovalenko,
  Phys.\ Lett.\  {\bf B352}, 1-7 (1995) 
  [hep-ph/9502315];
 M.~Hirsch, H.~V.~Klapdor-Kleingrothaus, S.~G.~Kovalenko,
  Phys.\ Rev.\  {\bf D53}, 1329-1348 (1996)
  [hep-ph/9502385];
   M.~Hirsch, H.~V.~Klapdor-Kleingrothaus, S.~G.~Kovalenko,
  Phys.\ Rev.\  {\bf D54}, 4207-4210 (1996)
  [hep-ph/9603213].
\bibitem{gerda}
 M.~Agostini {\it et al.}  [GERDA Collaboration],
  Phys.\ Rev.\ Lett.\  {\bf 111}, 122503 (2013)
  [arXiv:1307.4720 [nucl-ex]].
%
\bibitem{gerdafuture}
  A.~A.~Smolnikov [GERDA Collaboration],
  arXiv:0812.4194 [nucl-ex].
\bibitem{Dimopoulos:1988fr} 
  S.~Dimopoulos, R.~Esmailzadeh, L.~J.~Hall and G.~D.~Starkman,
  Phys.\ Rev.\ D {\bf 41}, 2099 (1990).

\bibitem{Hewett:1998fu} 
  J.~L.~Hewett and T.~G.~Rizzo,
  In *Vancouver 1998, High energy physics, vol. 2* 1698-1702
  [hep-ph/9809525].

\bibitem{Dreiner:1998gz} 
  H.~K.~Dreiner, P.~Richardson and M.~H.~Seymour,
  hep-ph/9903419.

\bibitem{Dreiner:2000qf} 
  H.~K.~Dreiner, P.~Richardson and M.~H.~Seymour,
  hep-ph/0001224.


\bibitem{Dreiner:2000vf} 
  H.~K.~Dreiner, P.~Richardson and M.~H.~Seymour,
  Phys.\ Rev.\ D {\bf 63}, 055008 (2001)
  [hep-ph/0007228].





\bibitem{Dreiner:2012np} 
  H.~K.~Dreiner and T.~Stefaniak,
  Phys.\ Rev.\ D {\bf 86}, 055010 (2012)
  [arXiv:1201.5014 [hep-ph]].

\bibitem{Allanach:2009xx} 
  B.~C.~Allanach, C.~H.~Kom and H.~Pas,
  JHEP {\bf 0910}, 026 (2009)
  [arXiv:0903.0347 [hep-ph]].

\bibitem{Allanach:2009iv} 
  B.~C.~Allanach, C.~H.~Kom and H.~Pas,
  Phys.\ Rev.\ Lett.\  {\bf 103}, 091801 (2009)
  [arXiv:0902.4697 [hep-ph]].
\bibitem{Reviews} 
  R.~Barbier, C.~Berat, M.~Besancon, M.~Chemtob, A.~Deandrea, E.~Dudas, P.~Fayet and S.~Lavignac {\it et al.},
  Phys.\ Rept.\  {\bf 420}, 1 (2005)
  [hep-ph/0406039].
\bibitem{Allanach:1999ic} 
  B.~C.~Allanach, A.~Dedes and H.~K.~Dreiner,
  Phys.\ Rev.\ D {\bf 60}, 075014 (1999)
  [hep-ph/9906209].
\bibitem{Bhattacharyya:2009hb} 
  G.~Bhattacharyya, K.~B.~Chatterjee and S.~Nandi,
  Nucl.\ Phys.\ B {\bf 831}, 344 (2010)
  [arXiv:0911.3811 [hep-ph]].
\bibitem{Aoyama:2012wj} 
  T.~Aoyama, M.~Hayakawa, T.~Kinoshita and M.~Nio,
  Phys.\ Rev.\ Lett.\  {\bf 109}, 111807 (2012)
  [arXiv:1205.5368 [hep-ph]].
\bibitem{Chatrchyan:2013qha} 
  S.~Chatrchyan {\it et al.}  [CMS Collaboration],
  Phys.\ Rev.\ D {\bf 87}, no. 11, 114015 (2013)
  [arXiv:1302.4794 [hep-ex]].


\bibitem{Belyaev:2012qa} 
  A.~Belyaev, N.~D.~Christensen and A.~Pukhov,
  Comput.\ Phys.\ Commun.\  {\bf 184}, 1729 (2013)
  [arXiv:1207.6082 [hep-ph]].

\bibitem{Sjostrand:2006za} 
  T.~Sjostrand, S.~Mrenna and P.~Z.~Skands,
  JHEP {\bf 0605}, 026 (2006)
  [hep-ph/0603175].


\bibitem {sarah} F.~Staub,
  arXiv:0806.0538 [hep-ph];
  Comput.\ Phys.\ Commun.\  {\bf 181}, 1077 (2010)
  [arXiv:0909.2863 [hep-ph]]; 
  Comput.\ Phys.\ Commun.\  {\bf 182}, 808 (2011)
  [arXiv:1002.0840 [hep-ph]].
 
   \bibitem {spheno} W.~Porod,
  Comput.\ Phys.\ Commun.\  {\bf 153}, 275 (2003)
  [hep-ph/0301101]; W.~Porod and F.~Staub,
  arXiv:1104.1573 [hep-ph].

\bibitem{Kretzer:2003it} 
  S.~Kretzer, H.~L.~Lai, F.~I.~Olness and W.~K.~Tung,
  Phys.\ Rev.\ D {\bf 69}, 114005 (2004)
  [hep-ph/0307022].


\bibitem{Chatrchyan:2011ds} 
  S.~Khatrchyan {\it et al.}  [CMS Collaboration],
  JINST {\bf 6}, P11002 (2011)
  [arXiv:1107.4277 [physics.ins-det]];
CMS Collaboration, CMS PAS EGM-10-004; 
CMS Collaboration, CMS PAS JME-10-014; 
CMS Collaboration, CMS PAS MUO-10-004.

\end{thebibliography}
\end{document}